\documentclass[12pt]{article}
\usepackage{latexsym,amsfonts}
\usepackage[round]{natbib}
\usepackage{float}
\usepackage{graphicx}

\newcommand{\be}{\begin{equation}}
\newcommand{\ee}{\end{equation}}
\newcommand{\bea}{\begin{eqnarray}}
\newcommand{\eea}{\end{eqnarray}}
\newcommand{\bean}{\begin{eqnarray*}}
\newcommand{\eean}{\end{eqnarray*}}

\newcommand{\halmos}{\vspace{3mm} \hfill \mbox{$\Box$}}
\newcommand{\R}{\mathbb{R}}

\newcommand{\sect}[1]{\section{#1} \setcounter{equation}{0}
                      \setcounter{table}{0} \setcounter{figure}{0}}

\newtheorem{defi}{Definition}[section]
\newtheorem{ex}[defi]{Example}

\newtheorem{thm}[defi]{Theorem}

\title{Ole E.\  Barndorff-Nielsen: Sand, Wind and Inference}

\author{Michael S\o rensen \\ 
{\small Department of Mathematical Sciences, University of Copenhagen,
  \vspace{-1.5mm}} \\ 
 {\small Universitetsparken 5, DK-2100 Copenhagen {\O}, Denmark}}

\begin{document}

\maketitle

\begin{abstract}
  \vspace{1mm}
This paper reviews Ole Eiler Barndorff-Nielsen's research in the first
decades of his career. The focus is on topics that he kept returning to
throughout his scientific life, and on papers that he built on in later
 important contributions. First his early contributions to the
foundations of statistical inference are reviewed with focus on
conditional inference and exponential families, two topics in which he
had a lifelong interest. The second half of the paper reviews his
research on wind blown sand and hyperbolic distributions and
processes, including his early contributions to modelling of turbulent
wind fields. This research laid the foundations for his later work on
financial econometrics and ambit processes.

\vspace{3mm}


\end{abstract}


\sect{Introduction}

Over a period of six decades, the Danish statistician and mathematician
Ole Eiler Barndorff-Nielsen (1935 - 2022) made vast and innovative
contributions in several fields of knowledge, including statistics,
probability, financial econometrics and theoretical physics. His early
research was mainly 
focused on the foundations of statistics, in particular sufficiency
and conditional inference and the exact theory of exponential
families. It is not possible to review all Barndorff-Nielsen's
early contributions in this area, many of which are included in his
book \cite{oebn78b}. In the first half of this paper, we shall
therefore mainly focus on work that played a significant role in his
later research, and in particular on contributions related to conditional
inference, a subject in which he had a standing interest during his
entire career. A brief presentation of main contributions to the
application of differential geometric methods in statistical
inferences is included - again with focus on relations to conditional
inference. His later work on approximate conditional inference, 
including the Barndorff-Nielsen formula for the conditional
distribution of the maximum likelihood estimator given an exact or
approximate ancillary statistic (originally called the $p^*$-formula,
\cite{oebn80b,oebn83b}), is reviewed in \cite{Battey} and will not be
presented here.

In the second half of the present paper, we review Barndorff-Nielsen's
main contributions to sand research and hyperbolic
distributions and processes, including his early contributions to
modelling of turbulent wind fields. The original motivation for
introducing the hyperbolic distribution was as a mathematical model
of the log-size distribution of sand grains in natural deposits, but
the distribution and its generalizations turned out to be widely
applicable, not least in financial econometrics and for modelling turbulence.
It is characteristic of Barndorff-Nielsen's research that he presented the
hyperbolic distribution as a member of the class of generalized
hyperbolic distributions, defined as normal variance-mean mixtures,
and proposed several stochastic process models with generalized hyperbolic
marginals, thus suggesting mechanisms that might explain the occurence
of hyperbolic distributions. The work on these distributions and
processes laid the foundations for his later research
contributions to financial econometrics and ambit processes, reviewed in
\cite{Shephard} and \cite{Veraart}, respectively.

Barndorff-Nielsen firmly believed that collaboration
with researchers from other fields is essential to inspire
theoretical statistical work and keeping it on track. He expected that
the physical sciences, including earth science, would require and
inspire development
of significant new directions in statistical science. Moreover, he
thought that statisticians should be general scientists with a
genuine interest in the fields for which they developed and applied
statistical methodology. That is certainly how he saw himself, as
transpires clearly from his autobiography {\it Stochastics in Science
  - Some autobiographical notes}, \cite{oebn18a}. He became so
deeply interested in the research on sand transport by wind that he
took the initiative to establish a group of earth scientists,
physicists and statisticians at the University of Aarhus, colloquially
called the ``Sand Gang'',
that jointly worked on sand transport, theoretically as well as
experimentally - both in the field and in a wind tunnel, for which he was
involved in obtaining funds. For several years Barndorff-Nielsen was
the inspiring leader of the gang and took eagerly part in all aspects
of the work - including field trips.

\sect{Early work on statistical inference}

While Barndorff-Nielsen studied actuarial science at the University of Copenhagen,
he worked part time as an assistant at the Biostatistics Department of
the Danish State Serum Institute.  This job was crucial for his development
as a scientist as the following quotes from the autobiography
\citet[pp.\ 16-18]{oebn18a} show. ``Due to the  exceptional research  
environment created by Michael Weis Bentzon and Georg Rasch, the
position was decisive for my whole future. ... There was, naturally, much routine
work to be done, but Bentzon and Rasch soon involved me in discussions
of research problems connected,  sometimes in a rather loose sense, to
the problems of actual importance to the responsibilities of the
Department. ... It was around this time Georg Rasch developed the
theory of measurement models which, in terms of applications, has had
a major impact in psychometrics and social science, and which
gave essential impetus to parts of the lively discussions on
conditional inference that took place from the mid-50s to the
mid-80s. ... This was the origin of my subsequent and standing
interest in conditional inference''.


\subsection{Conditional inference and parameters of interest}

Rasch's measurement model, \cite{rasch}, contained many more parameters than the
parameters of interest. These nuisance parameters caused problems,
which, as Rasch had noticed, could be avoided by conditioning on
certain statistics. This was the reason why conditional inference was such a
central issue in the discussions at the Danish State Serum Institute.

According to the conditionality principle, the relevant frame of reference, when 
drawing statistical inference, is all possible data sets with the same
value of an ancillary statistic as the given data set. The purpose of
conditioning on the ancillary statistic is to take into account the
quality, in a broad sense, of the given data set when drawing
conclusions. The principle begs the obvious question: when can a
statistic be said to be ancillary? This problem intrigued
Barndorff-Nielsen for many years, although he 
later became more interested in making the conditionality principle
operational. An ancillary statistic should not contain information about the
parameters of interest, but should indicate particular aspects of the
given data set that ought to be taken into account when drawing
conclusions about these parameters.

Barndorff-Nielsen formulated a framework for discussing the concepts
of sufficiency and ancillarity, see e.g.\ \citet[Part
1]{oebn78b}. Consider a statistical model $\mathcal{P} 
= \lbrace P_ \theta \, \vert \, \theta \in \Theta \rbrace$ for
observations $x \in {\cal X}$ of a random variable $X$ and a statistic
$T = t(X)$, where $t: {\cal X} \rightarrow {\cal Y}$. If the measures
in $\mathcal{P}$ are dominated by a probability measure, then 
\be
\label{gendecomp}
f_X (x;\theta) = f_T(t(x);\theta) f_{X|T}(x\,|\,t(x);\theta),
\ee
where $f_X(x;\theta)$, $f_T(t;\theta)$ and $f_{X|T}(x \,|\,t;\theta))$
are the density functions of $X$ and $T$
and of the conditional distribution of $X$ given $T=t$,
respectively. The statistic $T$ is sufficient in the classical sense
if $f_{X|T}$ does not depend on $\theta$, while it is ancillary in the
classical sense if $f_T$ does not depend on $\theta$. Barndorff-Nielsen
called these classical concepts B-sufficiency and B-ancillarity. Now
suppose $\theta = (\psi, \eta)$, where $\psi$ is the parameter of interest,
as is the situation for the Rasch model. A simpler example that illustrates
why it may, in such situations, be desirable to condition on a
statistic, is the well-known Neyman-Scott problem, see \cite{neymanscott}.

\begin{ex}
{\em Suppose
\[
X_ {ij} \sim N (\xi_{i}, \sigma ^ 2), \ \ \ i = 1, \dots, n, \ \ j = 1,2,
\]
are independent, and that  $\sigma^2$ is the parameter of
interest. The maximum likelihood estimators are $\hat \xi_i = \bar X _{i } =
\mbox{ \small $\frac{1}{2}$} (X_{i1} +X_{i2})$ and  ${\hat \sigma}^2 =
\mbox{ \small $\frac{1}{2n}$}  SSD$, where  $SSD = \sum _{ij}(X_{ij}
-{\bar {X}}_{i})^2$. However,  ${\hat \sigma}^2$ is strongly biased and 
inconsistent: $E( {\hat \sigma} ^ 2) = \frac{1}{2} \sigma^2$ and
$\hat \sigma ^2 \rightarrow \frac{1}{2} \sigma ^2$ almost surely as $n \rightarrow
\infty$.

The statistic $(SSD, \bar X _{1},\ldots, \bar X _{n})$ is minimal
sufficient, and $SSD$ and  $(\bar X _{1},\ldots, \bar X _{n})$ are
independent, so if it could be argued that $(\bar X _{1},\ldots, \bar
X _{n})$ is in some sense ancillary for $\sigma^2$, then inference
should be based on $SSD$. Since $SSD \sim 2\sigma^2 \chi^2_n$, the
maximum likelihood estimator based on $SSD$ is ${\tilde \sigma}^2 =
SSD/n = 2  \hat \sigma ^2$, which is clearly unbiased and
consistent. However, ${\bar {X}}_{i} \sim N(\xi_i,\sigma^2/2)$, so the
concept of B-ancillarity is not applicable. Neither is $SSD$ B-sufficient
for $\sigma^2$, which would also have suggested inference based on $SSD$.
}

\halmos
  
\end{ex}

\cite{fraser} extended the concept of B-sufficiency by the definition
that a statistic $T$ is sufficient for $\psi$ if the following is
satisfied: the distribution of $T$ is independent of $\eta$, the
conditional distribution given $T$ is independent of $\psi$, and if
$(\psi,\eta) \in \Psi \times H$, i.e.\ if
\be
\label{S}
f_X (x; \psi,\eta) = f_T(t(x);\psi) f_{X|T}(x\,|\,t(x);\eta),
\hspace{3mm} (\psi,\eta) \in \Psi \times H.
\ee
\cite{sverdrup} and \cite{sandved} proposed the corresponding ancillarity
concept by saying that $T$ is ancillary for $\eta$ if (\ref{S})
holds. Barndorff-Nielsen introduced the names {\it S-sufficiency} and
{\it S-ancillary} for these notions as well as the useful definition that
$T$ is called a {\it cut} if (\ref{S}) holds. The notions of S-sufficiency
and S-ancillarity solves nuisance parameter problems for many models,
but does not, for instance, solve the Neyman-Scott problem.

For group generated statistical models, \cite{barnard} proposed a
notion of sufficiency for the parameter of interest in the absence of
knowledge of the nuisance parameter. This concept was called G-sufficiency in
\cite{oebn76a}, where the corresponding concept of G-ancillarity was
introduced. Barndorff-Nielsen also introduced the notions of
M-sufficiency and M-ancillarity, see \cite{oebn73a, oebn76b}, which,
however, he later abandoned. It follows from a result in
\cite{oebn76a} that G-sufficiency entails M-sufficiency and similarly
for ancillarity. M-ancillarity is closely related to a scheme for
statistical inference, called plausibility inference, which
Barndorff-Nielsen proposed in a paper, \cite{oebn76a}, read to the
Royal Statistical Society. The approach parallels likelihood theory with
the likelihood function $L(\theta) = f_X(x;\theta)$ replaced by the
plausibility function $\Pi(\theta) =f_X(x;\theta)/\sup_y
f_X(y;\theta)$. He argued that likelihood and plausibility throw light
on different aspects of the evidence in the data $x$.

Many years later, in \cite{oebn99a}, the notions of L-sufficiency and
L-ancil\-larity, which he found satisfactory, were
presented. \cite{remon} had introduced a notion of L-sufficiency. 
By this definition, a statistic $t(X)$ is L-sufficient for $\psi$,
if the normed profile likelihood function for $\psi$,
$\tilde{l}(\psi;x) = \sup_\eta f_X(x;\psi,\eta)/\sup_{\theta} f_X(x;\theta)$,
depends on $x$ only through $t(x)$. Barndorff-Nielsen called R\'emon's
concept weak L-sufficiency and strengthened it by the definition that  a
statistic $T = t(X)$ is L-sufficient for $\psi$, if the distribution
of $T$ depends only on $\psi$, and there exist functions $g(t;\psi)$
and $h(x)$ such that
\be
\label{Lsuf}
\sup_\eta f_{X|T}(x\,|\,t(x); \psi, \eta) = g(t(x);\psi)h(x).
\ee
The corresponding L-ancillarity is defined by saying that a statistic $T =
t(X)$ is L-ancillary for $\psi$, if the conditional distribution 
of $X$ given $T$ depends only on $\psi$ and there exist functions $g(\psi)$
and $h(t)$ such that
\be
\label{Lanc}
\sup_\eta f_{T}(t; \psi, \eta) = g(\psi)h(t).
\ee
If $T$ is L-sufficient, then it can be argued as follows that
inference on $\psi$ should be based on the observation $T$ and its
distribution. The profile likelihood function is the basis for
inference about $\psi$, and when $T$ is L-sufficient, 
the profile likelihood function equals the product of $f_{T}(t;
\psi)$ and $g(t;\psi)$ ($h(x)$ is irrelevant). The second factor
contains the extra knowledge we obtain by observing $X$ given $T =
t$, but since it has the same value no matter what $x$ (such that
$t(x) = t$), we have observed, it does not contain any such extra
information about $\psi$ and should not be used for inference. It is this
second factor that is the cause of the trouble in the Neyman-Scott
problem. In a similar way, it can be argued that if $T$ is L-ancillary
for $\psi$, then inference about $\psi$ should be based on the
conditional distribution of $X$ given $T$. 

\setcounter{defi}{0}
\begin{ex}
{\em (continued) If we set $\psi = \sigma^2$ and $\eta = (\xi_1,
  \ldots, \xi_n)$, then it is easy to see that $(\bar X _{1},\ldots,
  \bar X _{n})$  is L-ancillary for $\psi$. The statistics $SSD$ and
  $(\bar X _{1},\ldots, \bar X _{n})$ are independent, and the
  distribution of $SSD$ does not depend on $\eta$. Hence the
  conditional distribution of $(SSD, \bar X _{1},\ldots,\bar X _{n})$
  given $(\bar X _{1},\ldots,\bar X _{n})$ is independent of
  $\eta$. Moreover,
  \[
\sup_ \eta f_{(\bar X _{1},\ldots, \bar X _{n})}
(x_1, \ldots, x_n; \sigma^2, \eta) =
\sup_ {\xi_1, \ldots, \xi_n}
\prod_{i=1}^n\frac{1}{\sqrt{\pi\sigma^2}}e^{-\sigma^{-2}(x_i-\xi_i)^2}
 = \left( \frac{1}{\pi\sigma^2}\right)^{n/2} .
\]    
Thus $(\bar X _{1},\ldots, \bar X _{n})$ is L-ancillary for
$\sigma^2$. Because $SSD$ and  $(\bar X _{1},\ldots, \bar X _{n})$
are independent, this also shows that $SSD$ is L-sufficient for $\sigma^2$.
}

\halmos
  
\end{ex}

Another interesting example is the well-known proportional hazards
models from survival analysis. The data are the waiting times until the
deaths of $n$ independent persons, and it is assumed that the hazard
function for the waiting time of $i$th person is $\lambda (t) \exp
\lbrace \beta \cdot z_{i}\rbrace$, 
where $\lambda$ is a positive function on $(0,\infty)$, $\beta \in
\mathbb {R} ^ {k}$, $ z_ {i} \in \mathbb {R} ^ {k}$ is a vector of
covariates for person no.\ $i$, and $\cdot$ denotes the inner
product. For this model it can be shown that the statistic 
$(R_1,\ldots,R_n)$ is L-sufficient for $\beta$, where $R_j \in \{ 1,
\ldots, n \}$ indicates who is the $j$th person to die. This suggests that
inference on $\beta$ should be based on the marginal distribution of
$(R_1,\ldots,R_n)$, which is Cox's partial likelihood function.

It is not difficult to show that S-sufficiency and S-ancillarity
implies L-sufficiency and L-ancillarity, respectively. It can also be
shown that G-sufficiency and G-ancillarity are particular cases of
L-sufficiency and L-ancillarity, which thus generalize and usefully
extend the most important notions of sufficiency and ancillarity
previously studied by Barndorff-Nielsen.  

\subsection{Exponential families}

A statistical model $\mathcal{P} = \lbrace P_ \theta \, \vert \, \theta \in
\Theta \rbrace$, $\Theta \subseteq \R^k$, for observations $x \in {\cal
  X}$ is called an exponential family if the measures in $\mathcal{P}$
are dominated by a $\sigma$-finite measure $\mu$ with densities of the
form 
\be
\label{expfam}
\frac{dP_\theta}{d\mu}(x) =  b(x)e^{\theta \cdot t(x) -
  \kappa(\theta)}, \ \ \theta \in \Theta,
\ee
where $t: {\cal X} \rightarrow \R^k$, $\kappa: \Theta \rightarrow
\R$, $b(x) \geq 0$ and $\theta \cdot t(x)$ denotes the inner product
of $\theta$ and $t(x)$. The statistic $t(X)$ is called a canonical
statistic. Here we use the so-called {\it canonical parametrization}, where
we assume that $\Theta$ is not contained in an affine subset of
$\R^k$. This parametrization is convenient for theoretical
considerations. Obviously, other parametrizations $\theta =
\psi(\alpha)$ could be used, and $b(x)$ could be absorbed in the
dominating measure $\mu$. Many widely used statistical models are
exponential families. A considerable part of Barndorff-Nielsen's
early research was devoted to exponential families, see the
comprehensive presentation in Part III of the book \cite{oebn78b}, 
and this model type continued to play an important role throughout his
research career.

A main contribution is the exact theory of maximum likelihood estimation for
exponential families, which applied convex analysis to clarify for which
observations $x$ a unique maximum likelihood estimator exists. Here exact
theory means a  finite sample theory as opposed to an approximate
asymptotic theory. In the following we will assume that the representation
(\ref{expfam}) is {\it minimal} in the sense that $k$ is the smallest
possible dimension of a canonical statistic $t$. This is the case if
and only if $t_1, \ldots, t_k$ and $1$ are linearly independent. Define
the convex set
\be
\label{domain}
\tilde{\Theta} = \left\{ \theta \in \R^k : \int e^{\theta \cdot t(x)}
    b(x)d\mu(x) < \infty  \right\}.
\ee
Then the exponential family $\mathcal{P} $ is called {\it full} if
$\Theta = \tilde{\Theta}$. It is called {\it regular} if it is full and the set
$\tilde{\Theta}$ is open. Note that the cumulant transform (i.e.\ the
logarithm of the Laplace transform) of the distribution of $t(X)$ under $P_\theta$ is
$\zeta \mapsto \kappa (\zeta + \theta) - \kappa (\theta)$ with domain
$\zeta \in \tilde \Theta - \theta$. Hence for $\theta \in $ int
$\tilde \Theta$ (the interior of $\tilde \Theta$), the expectation of $t(X)$
under $P_\theta$ is given by
\be
\label{mean}
\tau(\theta) := E_\theta (t(X)) = \nabla \kappa (\theta),
\ee
where $\nabla \kappa$ denotes the gradient of $\kappa$. The function
$\tau$ maps int $\! \tilde \Theta$ into
the closed convex hull of the support of the distribution of $t(X)$,
which does not depend on $\theta$. Note also that the likelihood
equation is $\tau(\theta) = t(x)$, and that the Fisher 
information matrix as well as the observed information matrix equals
$-H \kappa$, where $H \kappa$ is the Hessian matrix of $\kappa$. The
convex function $\kappa$ is called steep if
\[
\lim_{\lambda \downarrow 0}\, (\theta^* -\theta)\cdot \nabla \kappa
(\lambda \theta + (1-\lambda)\theta^*) = \infty 
\] 
for every $\theta \in$ int $\! \tilde \Theta$ and every $\theta^*$ on the
boundary of $\tilde \Theta$. A full exponential family for which
$\kappa$ is steep is call steep. A regular exponential family is steep.

Barndorff-Nielsen's main result on maximum likelihood estimation
for exponential families is the following theorem.
\begin{thm}
  \label{theoremmle}
Suppose the exponential family $\mathcal{P} $ is full and that the
representation (\ref{expfam}) is minimal. Then the
maximum likelihood estimator exists if and only if  $t(x) \in $ {\em int}
$\!  C$, where $C$ denotes the closed convex hull of the support of the
distribution of $t(X)$. If the maximum likelihood estimator exists, it
is unique. The mapping $\tau$ is injective, and $\tau(${\em int} $\!
\Theta) \subseteq$ {\em int} $\! C$. If $t(x) \in \tau 
(\mbox{\em int } \! \Theta)$, then the maximum likelihood estimator $\hat
\theta$ is the solution to the likelihood equation, $\hat \theta =
\tau^{-1}(t(x))$.

If the exponential family is, moreover, steep (in particular regular), then $\tau$
is a bijection of {\em int} $\! \Theta$ onto {\em int} $\!  C$, so for $t(x)
\in $ {\em int} $\!  C$ the maximum likelihood estimator is always given
by $\hat \theta = \tau^{-1}(t(x))$.
\end{thm}  

An alternative parametrization of a full exponential family is the
{\it mean value parametrization}  $\tau = \tau (\theta)$, for which
the maximum likelihood estimator is $\hat \tau = t(x)$ when $t(x)
\in \tau (\mbox{int } \! \Theta)$.  Barndorff-Nielsen introduced the useful {\it
  mixed parametrization}. Consider a partition of the canonical
statistic given by $t(x) = (t^{(1)}(x), t^{(2)}(x))$ and the
corresponding partitions of the canonical parameter $\theta =
(\theta^{(1)}, \theta^{(2})$ and the mean value parameter
$(\tau^{(1)}, \tau^{(2)})$, \ where $\tau^{(i)} =
E_{\theta}(t^{(i)}(X))$, $i=1,2$. It can be shown, see \citet[pp.\
121-122]{oebn78b},  that the mapping 
$\theta \mapsto (\tau^{(1)}, \theta^{(2)})$ is a homeomorphism, and
that hence, in particular,  $(\tau^{(1)}, \theta^{(2)})$ is a
parametrization of $\mathcal{P}$. For a regular exponential family,
the components $(\tau^{(1)}$ and $\theta^{(2)})$ are variation
independent, i.e.\ the parameter set is a product set corresponding to
the partition.

The concept of S-ancillarity in exponential families was studied
thoroughly in \cite{oebn75a}. As explained in the previous subsection,
the existence of an S-sufficient statistic or an S-ancillary
statistic is equivalent to the existence of a cut, i.e.\ a statistic such
that (\ref{S}) holds. A main result is the following theorem, which is
concerned with a partion $t(x) = (t^{(1)}(x), t^{(2)}(x))$ of the
canonical statistic and the corresponding mixed parametrization
$(\tau^{(1)}, \theta^{(2)})$.

\begin{thm}
\label{theoremS}
Suppose $\Theta$ is an open and connected set. Then $t^{(1)}(X)$ is a
cut if and only if

\begin{description}

\item[(i)] $\tau^{(1)}$ and $\theta^{(2)}$ are variation independent

\item[(ii)] there exist functions $\phi$ and $\chi$ such that
  $\theta^{(1)} = \phi (\tau^{(1)}) + \chi (\theta^{(2)})$.

\end{description}
If the exponential family is regular, then condition (ii) is
sufficient (because (i) is automatically satisfied).
\end{thm}

When the statistic $t^{(1)}(X)$ is a cut, then it is S-ancillary for
$\theta^{(2)}$ and S-sufficient for $\tau^{(1)}$. Cuts in exponential
families were discussed further in \cite{oebn76c}, where a method for
constructing regular exponential families with cuts was proposed. Cuts
in natural exponential families, i.e.\ families where $t(x) = x$, were
treated in \cite{oebn95a}.

Among other significant contributions to the theory of exponential families
is the investigation of exponential transformation models in
\cite{oebn82e}. The topic is models in the intersection between the
two well-studied general classes of statistical models, both
originally introduced by R.A. Fisher, the
exponential families and the transformation models. A transformation
model is a statistical model for observations $x \in {\cal X}$ of the
form $\{ h(P_0) \, | \, h \in H \}$, where $P_0$ is a probability
measure on $\cal X$, $H$ is a group of one-to-one transformations of
$\cal X$ onto itself, and $h(P_0)$ denotes the transformation of the
measure $P_0$ by $h$. The paper studies the interaction between the
exponential form of the density and the group action, discusses the
structure of inference 
for exponential transformation models, and investigates extensions such as
the full exponential family generated by the original transformation
model, which is an orbit in the full model.

Another contribution worth mentioning is the study of reproductive
exponential families in \cite{oebn83a}. These models have a property
that generalizes the ``closed under convolution'' property of the classes
of normal distributions, gamma distributions and inverse Gaussian
distribution. The fact that the last two classes of distributions are
closed under convolution gives them a significant role in
Barndorff-Nielsen's theory of the generalized hyperbolic distributions
and processes discussed in the second part of this paper, which he later applied to
model turbulence and financial data.

\subsection{Differential geometry and statistical inference}

Fascinated by Shun'ichi Amari's research on information geometry,
Barndorff-Nielsen became involved in the investigation of what
insights might be gained by applying differential geometric methods to
statistical models considered as differentiable manifolds. In the
differential geometric approach to parametric inference, developed by
Chensov, Efron, Amari and others, the parameter space is set up as a
differentiable manifold equipped with a Riemannian metric where the
metric tensor is the Fisher information matrix. A family of affine
connections is determined by the Fisher information matrix and the
so-called expected skewness tensor of the score vector. Drawing on the
ideas of ancillarity and conditional inference, Barndorff-Nielsen made
his most significant contribution to the differential geometric
approach in the paper \cite{oebn86a}, where he proposed to replace the
expected geometries used in previous work by observed geometries that
are data dependent. His motivation for doing this was to throw light on
likelihood inference and asymptotic expansions. Whereas the expected
geometries give useful geometric interpretations of Edgeworth
expansions, the observed geometries are tailored to approximations of
conditional distributions like Barndorff-Nielsen's $p^*$-formula, see
\cite{Battey}. 

Consider a statistical model $\mathcal{P} = \lbrace P_ \theta \, \vert \, \theta \in
\Theta \rbrace$, $\Theta \subseteq \R^k$, for observations $x \in {\cal
  X}$ with log likelihood function $\ell (\theta;x)$. Assume that
the maximum likelihood estimator $\hat  \theta$ exists, and that an
auxiliary statistic $a = a(x)$ exists such that $(\hat \theta, a)$ is
minimal sufficient for $\mathcal {P}$. The observed information is
given by  $j(\theta;x) = - H \ell (\theta;x)$, where $H  \ell$ denotes
the Hessian matrix of the log likelihood function. Since the
likelihood function depends on $x$ only through the sufficient
statistic $(\hat \theta, a)$, we can write the log likelihood function
and the observed information in the form
\[
\ell (\theta; \hat \theta, a) \ \ \mbox{ and } \ \ j (\theta; \hat \theta, a). 
\]
The metric tensor of the observed geometry on $\Theta$ is given by $j
\hspace{-2.2mm} \backslash (\theta; a) = j (\theta; \theta, a)$. The
family of affine connections is determined by $j \hspace{-2.2mm}
\backslash (\theta; a)$ and observed skewness tensor $T \hspace{-2.5mm} 
\backslash (\theta; a)= T(\theta; \theta, a)$,  where the $ijk$'th
element of $T$ is given by  
\bea
\label{mixed}
\lefteqn{T_{ijk}(\theta;\hat \theta, a) =}  \\ && - \left( 
\partial_{\theta_i} \partial_{\theta_j} \partial_{\theta_k}
\ell(\theta; \hat \theta, a) +  
\partial_{\theta_i} \partial_{\theta_j} \partial_{\hat \theta_k}
\ell(\theta; \hat \theta, a) +  
\partial_{\theta_j} \partial_{\theta_k} \partial_{\hat \theta_i}
\ell(\theta; \hat \theta, a) +  
\partial_{\theta_k} \partial_{\theta_i} \partial_{\hat \theta_j}
\ell(\theta; \hat \theta, a)   \right).  \nonumber
\eea
In the observed geometry mixed derivatives of the log likelihood
function (differentiation with respect to both $\theta$ and $\hat
\theta$ as in (\ref{mixed})) take the place of the moments of derivaties
of the log likelihood function that appear in the expected
geometry. The observed geometry depends on the data through the
statistic $a = a(x)$. Estimation and inference obtained via the observed
geometry will hence be conditional on $a$. In the construction of the
geometry it is not necessary to assume that 
the statistic $a$ is ancillary, but when applying the geometry to
statistical inference, it is essential to choose $a$ such that it is
in some sense ancillary, at least to the relevant asymptotic order.

Suppose $\theta = (\psi, \eta)$, where $\psi \in \Psi$ is a
parameter of interest. Statistical geometries on the manifolds
$\Theta$ on $\Psi$ as well as transfer of geometries from $\Theta$ to $\Psi$
were investigated in \cite{oebn88b}, which includes a thoroughly
survey of transfer of geometries along submersions (e.g.\ projections
$\R^{k+m} \mapsto \R^k$).  A main result concerns the transfer of the
observed geometry on $\Theta$ to $\Psi$ and the 
relation of the transferred geometry to the observed profile geometry
on $\Psi$ given by $(\tilde j \hspace{-2.2mm} \backslash, \tilde T
\hspace{-2.5mm} \backslash )$ defined by the same construction as
above, but using the profile log-likelihood 
\[
 \tilde \ell (\psi; x) = \sup_ {\eta} \ell (\psi,\eta;x) 
\]
instead of $\ell (\theta; x)$. Interestingly, the concept of
L-sufficiency is needed here.
\begin{thm}
\label{theoremGeom}
Assume that the statistic U = u(X), where $u: {\cal X} \mapsto {\cal
  Y}$, is L-sufficient for $\psi$, and that the auxiliary functions $a:  {\cal
  X} \mapsto A$ and $b:  {\cal Y} \mapsto B$ are such that
\begin{description}

\item[(i)] $(\hat \theta, a) :  {\cal X} \mapsto \Theta \times A$ is
  bijective and $(\hat \theta, a(X))$ is sufficient

\item[(ii)] $(\hat \psi, b) : {\cal Y} \mapsto \Psi \times B$ is a bijection

\item[(iii)]  $b(u(x))$ depends on $x$ only through $a(x)$.

\end{description}
Then $(j \hspace{-2.2mm} \backslash, T \hspace{-2.5mm}  
\backslash )$ transfers into the observed profile geometry
$(\tilde j \hspace{-2.2mm} \backslash, \tilde T \hspace{-2.5mm}
\backslash ).$
\end{thm}
When an L-sufficient statistic for $\psi$ exists, also an expected
profile geometry can be defined on $\Psi$, but it is not known under
which conditions the expected geometry on $\Theta$ transfers to
$\Psi$. An example exists where it does transfer, but not into
the expected profile geometry.

A review of Barndorff-Nielsen's approach to applying differential
geometric methods to statistical inference can be found in
\cite{oebn85c}. He also published a series of papers that developed
mathematical concepts and theory to facilitate the use of differential
geometry in statistics, see e.g.\ \cite{oebn87a}.

\sect{Hyperbolic distributions, sand and turbulence}

Around 1969 Ole Barndorff-Nielsen was approached by Jens Tyge M{\o}ller, who
was a professor at the Department of Earth Sciences at Aarhus
University. He was worried that there might be problems with the laboratory
procedures at the department. In the geomorphology literature it was the
traditional consensus that the distribution of the logarithm of the
grain size in deposits of wind blown sand was Gaussian. However, for
sand samples from the Danish West Coast, Jens Tyge M{\o}ller and his
colleagues invariably found tails that were heavier than Gaussian
tails. Importantly, M{\o}ller then asked Barndorff-Nielsen whether he
could make sense of a log-log plot of the density of the empirical size
distribution of a sand sample from the Libian Desert on page 155 in
Ralph Alger Bagnold's book {\it The Physics of Blown Sand and Desert
  Dunes}, \cite{bagnold}, which was similar to the Danish size
distributions. The plot strongly indicated a hyperbolic shape rather
than the Gaussian parabolic shape, but this finding had not been taken
seriously by geomorphologists.

\subsection{Hyperbolic distributions}

Barndorff-Nielsen was intrigued and formalized Bagnold's heuristic
ideas mathematically by introducing, in \cite{oebn77a}, the {\it hyperbolic distribution},
$H(\alpha, \beta, \delta, \mu)$, with density function
\be
\label{hyp}
\frac{\gamma}{2\alpha\delta K_1(\delta\gamma)}\exp{\left\{-\alpha
  \sqrt{\delta^2+ (x-\mu)^2} + \beta(x-\mu) \right\}}, \ \ x \in \R
\ee
where $K_\lambda$ is a Bessel function, and $\alpha > |\beta| \geq 0, \
\delta > 0, \ \mu \in \R$ and $\gamma = \sqrt{\alpha^2 -
\beta^2}$. By construction, the logarithm of this density function is a hyperbola.
Importantly, he showed that the hyperbolic distribution is a normal variance-mean
mixture in the following sense. Let $X$ and $W$ be random variables
such that
\be
\label{mix}
X \, | \, W=w \, \sim \, N(\mu+\beta w, w),
\ee
and $W$ has the density function
\be
\label{gighyp}
\frac{\gamma/\delta}{2 K_1(\delta\gamma)}
\exp\left(-\mbox{\small$\frac12$}(\delta^2w^{-1}+ \gamma^2 w)\right),
\ \ w>0, \ \ \ \gamma>0, \ \delta >0.
\ee
Then $X \sim H(\alpha, \beta, \delta, \mu)$ with $\alpha = \sqrt{\beta^2 +
\gamma^2}$.

Barndorff-Nielsen noted that the probability distribution given by
(\ref{gighyp}) belongs to the class of what he called {\it generalized
inverse Gaussian distributions} ($GIG$). The $GIG (\lambda, \delta,
\gamma)$-distribution has density function
\be
\label{gig}
\frac{(\gamma/\delta)^{\lambda}}{2 K_{\lambda}(\delta\gamma)} w^{\lambda - 1}
\exp\left(-\mbox{\small$\frac12$}(\delta^2w^{-1}+ \gamma^2 w)\right),
\ \ w>0, 
\ee
where $\lambda \in \R$, and where $\delta>0, \gamma \geq 0 $ if
$\lambda<0$, $\delta>0, \gamma>0$ if $\lambda=0$ and $\delta\geq 0,
\gamma>0$ if $\lambda>0$. For $\lambda = -1/2$ this is the inverse
Gaussian distribution, hence the name of the class. The gamma
distribution is obtained for $\lambda > 0, \gamma > 0$ and $\delta=0$.
The class of distributions with density function (\ref{gig}) was first proposed 
in the 1940s by the French statistician \'Etienne Halphen to fit data  on the
monthly flow of water in hydroelectric stations, see \cite{Halphen}. The class
was rediscovered by \cite{sichel} and Barndorff-Nielsen, and was
briefly mentioned by \cite{good} as an intermediate between the Pearson 
distributions of Type III and V.

Using this generalization of (\ref{gighyp}), Barndorff-Nielsen
defined the {\it generalized hyperbolic distribution} 
$GH(\lambda,$ $\alpha, \beta, \delta, \mu)$  as the normal variance-mean
mixture obtained when the distribution of $W$ in (\ref{mix}) is the $GIG
(\lambda, \delta, \gamma)$-distribution with $\gamma = \sqrt{\alpha^2
  -\beta^2}$. The density function of the $GH(\lambda, \alpha, \beta,
\delta, \mu)$-distribution is  
\be
\label{gh}
\frac{(\gamma/\delta)^{\lambda}}{\sqrt{2\pi}K_{\lambda}(\delta\gamma)} \cdot 
  \frac{K_{\lambda-\mbox{\small{$\frac{1}{2}$}}}\left( \alpha\sqrt{\delta^2+
  (x-\mu)^2} \; \right)}{\left(\sqrt{\delta^2+(x-\mu)^2}/\alpha
  \right)^{\mbox{\small
  {$\frac{1}{2}$}}-\lambda}}\cdot e^{\beta(x-\mu)}, \ \ \ x \in \R
\ee
where $\lambda, \mu \in \R$, and where $\delta \geq 0, \alpha>
|\beta|$ if $\lambda>0$, $\delta>0, \alpha > |\beta|$ if 
$\lambda=0$ and $\delta>0, \alpha \geq |\beta|$ if  $\lambda<0$.

An important subclass is obtained for $\lambda = -1/2$, where the mixing
distribution is the inverse Gaussian distribution. Barndorff-Nielsen
called this subclass the {\it normal inverse Gaussian distributions}
($NIG$). In his autobiography, \cite{oebn18a} p.\ 142, he wrote that ``the
normal inverse Gaussian law is the member of leading interest in the
class of generalized hyperbolic distributions''. The density function of
the $NIG (\alpha, \beta, \delta, \mu)$-distribution is
\be
\label{nig}
 \frac{\alpha\delta}{\pi}e^{\delta\gamma}\cdot\frac{K_1 \left(\alpha
  \sqrt{\delta^2+ (x-\mu)^2} \right)}{\sqrt{\delta^2+(x-\mu)^2}}
  \cdot e^{\beta(x-\mu)}, \ \ \ x \in \R
\ee
where $\alpha \geq |\beta|, \delta > 0, \mu \in \R$. The
$NIG$-distributions are particularly useful because the class is
closed under convolution when $\alpha$ and $\beta$ are fixed:
\be
\label{nigconvolution}
NIG (\alpha, \beta, \delta_1, \mu_1) * NIG (\alpha, \beta, \delta_2,
\mu_2) = NIG (\alpha, \beta, \delta_1 + \delta_2, \mu_1 + \mu_2) 
\ee 
where $*$ denotes convolution, see \cite{oebn78a}. It has turned out
to be applicable in a wide range of fields, not least turbulence and
financial econometrics. 

Only two subclasses of the $GH$-distributions are closed under
convolution. The other class is 
obtained when the mixing distribution is a Gamma distribution
($\delta = 0, \lambda > 0, \gamma >0)$. \cite{sichel} introduced this
class of distributions to model the log-size-distribution of
diamonds. These distributions were later rediscovered in the
finance literature as models of financial returns by
\cite{madan&seneta}, who called them {\it variance gamma
  distributions} ($VG$). The density function of the $VG (\lambda,
\alpha, \beta, \mu)$-distribution is
\be
\label{vargamdens}
\frac{\gamma^{2 \lambda}}{\sqrt{\pi} \Gamma (\lambda) (2 \alpha)^{\lambda -
\frac12}} | x - \mu |^{\lambda - \frac12 } K_{\lambda -\frac12} \left( \alpha 
|x - \mu| \right) e^{\beta (x - \mu)},\;\;\;x\in\R,    
\ee
where $\Gamma$ denotes the gamma-function. The class is closed under
convolution when $\alpha$ and $\beta$ are fixed:
\be
\label{vgconvolution}
VG(\lambda_1, \alpha, \beta, \mu_1) *VG(\lambda_2, \alpha, \beta,
\mu_2) =  VG(\lambda_1 + \lambda_2, \alpha, \beta, \mu_1 +\mu_2 ),
\ee
see \cite{oebn78a}. Using (\ref{mix}), it is an easy exercise to show that the
convolution results (\ref{nigconvolution}) and (\ref {vgconvolution})
follow from the well-known convolution results for the normal inverse
Gaussian distribution and the gamma distribution.

The subclass $GL (-\nu, |\beta|, \beta, \delta,\mu)$ ($\nu,
\delta > 0$, $\beta, \mu \in \R$), obtained when the mixing
distribution is a reciprocal Gamma distribution, can be interpreted as
an {\it asymmetric scaled t-distribution}. When $\beta = 0$, it is the
t-distribution with $\nu$ degrees of freedom.

A final point in \cite{oebn77a} is the introduction of the class of
{\it multivariate generalized hyperbolic distributions} by considering in
(\ref{mix}) a $n$-dimensional normal distribution, i.e. $ X \, | \,
W=w \, \sim \, N_n(\mu+w \beta' \Delta, w \Delta)$, where $W \sim GIG
(\lambda, \delta, \gamma)$, $\mu, \beta \in \R^n$ and $\Delta$ is a
positive definite $n \times n$-matrix. In this way, the density
function
\be
\label{multivar}
\frac{1}{(2 \pi)^{n/2}\sqrt{|\Delta|}}
  \frac{(\gamma/\delta)^{\lambda}}{K_{\lambda}(\delta\gamma)} \cdot 
  \frac{K_{\lambda-\mbox{\small{$\frac{n}{2}$}}}\left( \alpha\sqrt{\delta^2+
  (x-\mu)'\Delta^{-1} (x-\mu)} \; \right)}{\left(\sqrt{\delta^2+(x-\mu)'\Delta^{-1} (x-\mu)}/\alpha
  \right)^{\mbox{\small
  {$\frac{n}{2}$}}-\lambda}}\cdot e^{\beta'(x-\mu)}, \ \ \ x \in \R^n,
\ee
with $\alpha = \sqrt{\beta' \Delta \beta + \gamma^2}$, is
obtained. The $n$-dimensional hyperbolic distribution is obtained
for $\lambda = (n+1)/2$. A more detailed introduction to the
generalized hyperbolic distributions can be found in
\cite{bibbysorensen}.

After the completion of the first paper on the hyperbolic
distributions, Barndorff-Nielsen contacted Bagnold, who was very
pleased that after 30 years someone had taken up his old suggestion of
the importance of the hyperbolic shape. This was the beginning of a
``delightful collaboration and friendship'' (\citet[p.\ 51]{oebn18a}), and soon they
wrote a joint paper, \cite{oebn80a}, with the title {\it The pattern
  of  natural size distributions}. In this and other papers it was
demonstrated that the hyperbolic distributions provide excellent fits
to the distribution of the log-size of samples from river bed
sediments, glacio-fluvial sediments, marine sediments, sand grains
that were jumping above the sand bed, diamonds in a mining area, and
droplets in aerosols, of the log-intensity of the remnant
magnetization of lava flows, of the log-length of words in 7 languages
(including Anglo-Saxon and Sanskrit), of personal income,
of the weight of beans, and of the difference between the velocities at
two points in space or time in atmospheric turbulence, see also
\cite{oebn83c} and the references given there.

In classical statistical mechanics, the distribution of the momentum of a
single gas particle in an ideal gas is a three-dimensional isotropic
Gaussian distribution, but if 
relativistic effects are taken into account the distribution is a
three-dimensional hyperbolic distribution, (\ref{multivar}) with $n=3$
and $\lambda = 2$, which is in general not
isotropic. A brief account of the general result, first established by
\cite{moller}, was given in \cite{oebn82a}, see also \citet[p.\ 65]{oebn18a}. 

It was pointed out in \cite{oebn78a} that the subclass of
$(n-1)$-dimensional generalized hyperbolic distributions with $\lambda =
n/2-1$ can be interpreted in a natural way as distributions on the
hyperboloid in $\R^n$ with properties analogous to the von Mises-Fisher
distributions on the sphere. These hyperboloid distributions were
studied thoroughly by \cite{jensen}. In particular, the statistical
properties were studied by an elegant group theoretic approach and
shown to be analogous to the statistical properties of the von
Mises-Fisher distributions. In fact, the setup revealed that the
mathematics behind the two classes of distributions is almost the
same. For $n=3$ the hyperboloid distributions provide a model for a
simultaneous record of a linear and an angular variable, which was
shown to give a good fit to simultaneous records of wind speed and
wind direction.

There are other classes of distributions with a similar ``hyperbolic''
shape. This was studied generally in \cite{oebn82c}, where conditions
on the mixing distribution were given which ensure that a normal
variance-mean mixture has hyperbolic tail behaviour. As a particular
example, the class of {\it generalized logistic distributions} (called
z-distributions in the paper) with density function
\be
\label{genlog}
\frac{1}{\sigma B(\alpha,\beta)} \frac{\exp ((x-\mu)/\sigma)^{\alpha}}{(1
  + \exp ((x-\mu)/\sigma)))^{\alpha+\beta}}, \ \ x \in \R, \hspace{3mm}
\alpha, \beta, \mu \in \R, \sigma > 0
\ee
was studied in detail and shown to have several mathematical properties in
common with the generalized hyperbolic distributions. In particular,
the distributions are normal variance-mean mixtures.

\begin{figure}
\begin{center}
\includegraphics[scale=0.4]{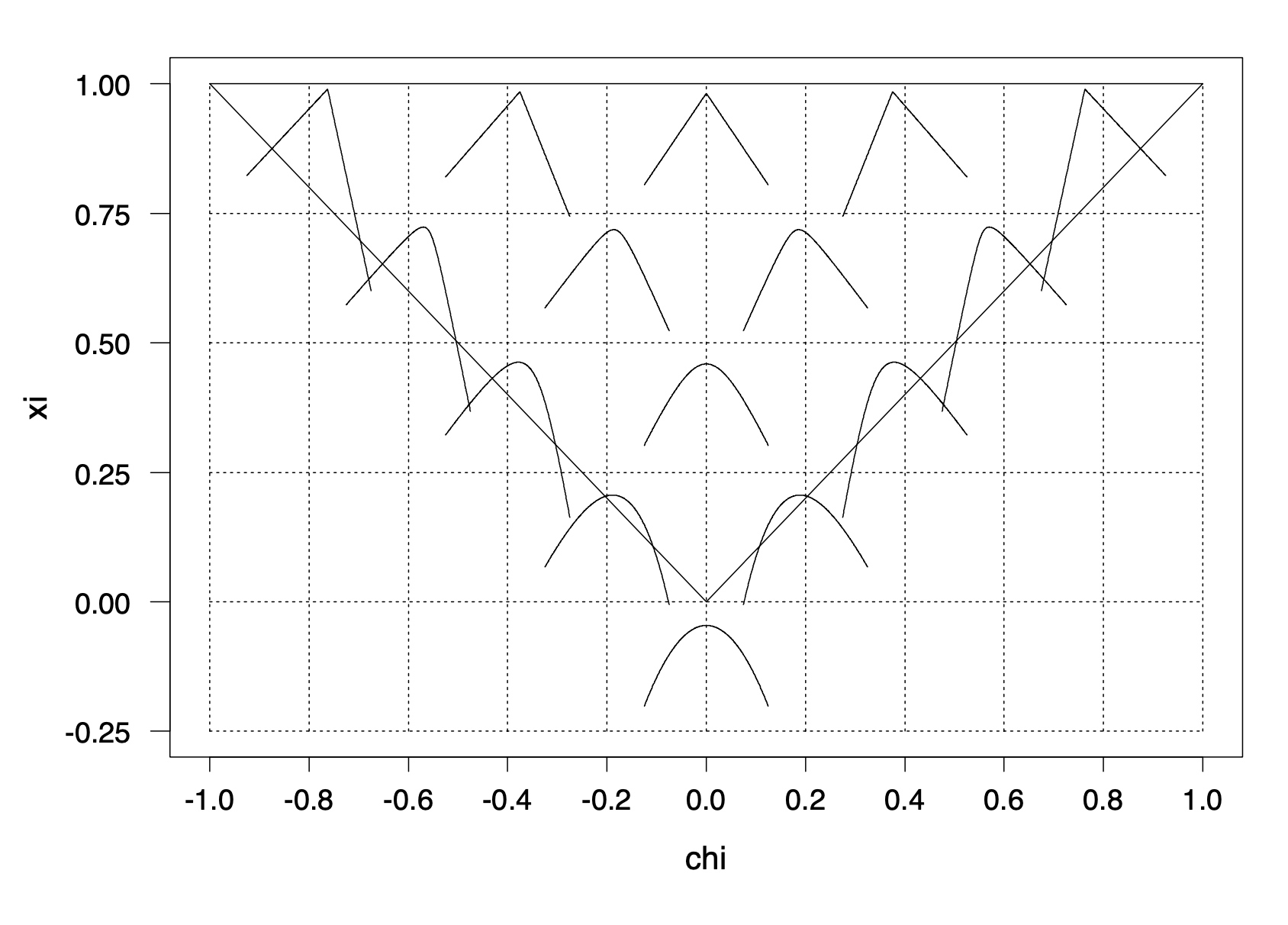} \vspace{-5mm}
\caption{\label{trianglefig} 
The hyperbolic shape triangle, with hyperbolic log-densities located below
their $(\chi,\xi)$-values.} 
\end{center}
\end{figure}

To facilitate the use of the hyperbolic distributions in statistical
practice, specifically to study how the variation differs
from Gaussian variation, the so-called
{\it hyperbolic shape triangle} was introduced in  \cite{oebn85a}. The
classical measures of non-normality, skewness and kurtosis, are rather
unwieldy expression involving Bessel functions. It was
therefore proposed use instead $(\chi,\xi)$ defined by
\[
\chi = \frac{\beta/\alpha}{\sqrt{1+\delta\gamma}} \hspace{10mm}
\xi = \frac{1}{\sqrt{1+\delta\gamma}},
\]
where $\gamma = \sqrt{\alpha^2 - \beta^2}$. For not too large values
of $|\beta|/\alpha$, the skewness and kurtosis are
roughly approximated by $(3 \chi, 3 \xi^2)$. The joint domain of
variation of $(\chi,\xi)$ is the triangle
\be
\label{shapetriangle}  
\left\{ (\chi, \xi) \in \R \, | \, 0 \leq |\chi | < \xi < 1  \right\},
\ee    
which makes it easy to detect patterns in estimates of $(\chi,\xi)$
from several samples. The shape triangle is shown in Figure
\ref{trianglefig}, where log-densities of hyperbolic distributions are
plotted for a number of $(\chi,\xi)$-values. The normal distribution
is located at $(0,0)$, while (mostly) skew Laplace distributions are
located at the upper boundary of the triangle ($\xi = 1$). 
An analogous  shape triangle has also been proposed for the class of
NIG-distributions, see \cite{rydberg}, and has been used to study how
the distribution of velocity differences in homogeneous turbulence
vary with different time lags or Reynolds numbers, see \cite{oebn04a}.

\subsection{Sand}

Barndorff-Nielsen took a deep
interest in research on sand transport by wind and established the
Sand Gang of earth scientists, statisticians and physicists at the
University of Aarhus. For a number of years, a considerable part of his time
was spent as the inspiring leader and driving force of the Sand
Gang. I joined the gang as a student and can attest to the fact that it was
an exciting and highly stimulating research environment with close
contact to the internationally leading research
environments in the field. Perhaps some impression of the atmosphere
is conveyed in the paper \cite{oebn85a}, see also part IV of
\cite{oebn18a}. In 1985 a highly succesful {\it International Workshop on 
the Physics of Blown Sand} was held at Aarhus University with
participation of most of the internationally leading researchers,
including Bagnold. The workshop was a decisive step in establishing an
international research community in this very interdisciplinary
area. A similarly succesful workshop was held in at Sandbjerg Manor,
also in Denmark, which consolidated and expanded the community. Today
these two workshops are considered the first in a series of
conferences that is still running under the name {\it International
  Conferences on Aeolian Research}, the most recent one was held in
2023 in Las Cruces, USA.

Inspired by Kolmogorov, who participated actively in empirical studies
of turbulence on board  a research vessel in the Pacific Ocean,
Barndorff-Nielsen also took a very active part in field  
studies and experiments, mostly carried out in sandy areas near the
West Coast of Denmark. During an early field study, reported in
\cite{oebn82b}, sand samples were collected from the
surface of a small barchanoid dune along a line in the wind direction
and were analysed by means of the hyperbolic distribution. In a more
thorough study of the data, \cite{oebn88a} were able to discriminate between
samples from different well-defined sections of the dune via their
positions in the shape triangle (\ref{shapetriangle}). Based on this
finding, a mathematical model for sand sorting was proposed. First the
authors introduced power-law erosion, where after a period of erosion
of length $t$, the proportion of grains of size $s$ still present at a
given location is proportional to $s^{\epsilon t}$, $\epsilon > 0$. If
the distribution of the log-size before the erosion period is
hyperbolic, i.e.\ has density function (\ref{hyp}), then at time $t$ the
log-size distribution is hyperbolic with the same parameter values,
except that $\beta (t) = \beta + \epsilon t$. Next they posited that a
negative value of $\epsilon$ can be interpreted as power-law
deposition, but also pointed out that deposition is presumably not
simply erosion in reverse. Power-law deposition does not take account
of the influence of the difference in size between the sand
particles. The authors argued that both grains that are small and
grains that are large relative to the most frequent grain sizes
(around the location parameter $\mu$) have a higher probability of
being deposited, and that deposition will change the grain size
composition of the sand bed, which in turn reflects back on the
deposition process. They formulated these considerations in
differential equations for the hyperbolic parameters $\alpha$ and
$\beta$
\[
\alpha'(t) = - \kappa, \hspace{6mm} 
\beta'(t) = - \left(\epsilon + \kappa  \frac{\beta(t)}{\alpha(t)}\right) 
\]
with the solution
\be
\label{spacecurve}
\alpha(t) = \alpha(0) - \kappa t, \hspace{5mm}
\kappa \frac{\beta(t)}{\alpha(t)} -\epsilon \log \alpha(t) = c_0,
\ee
where $\kappa/\epsilon < 0$ and $c_0$ is determined by the initial
conditions. Thus  
it is assumed that the distribution of the log-size is hyperbolic at
all times, where (\ref{spacecurve}) defines curves in the hyperbolic shape
triangle that show the effect of erosion and deposition and can be
compared to empirical $(\chi,\xi)$-values such as those from the
barchanoid dune mentioned above. The model was generalized to a
space-time model of sand sorting in \cite{oebn91a}.   

Very early on it was decided to built a wind tunnel at Aarhus
University to study the physics of drifting sand. Funds for this were
scarce, but through the tireless efforts of Barndorff-Nielsen and the
Sand Gang, a first version of a wind tunnel was built. The motor was
a second hand Rolls Royce jet engine donated by the Danish Air Force
after it had served its time in the air. The wind tunnel is still in
use and has over the years been improved in many ways and
supplemented by a wind tunnel that can simulate the atmospheric
conditions on Mars. Today sand researcher from all over the globe come
to Aarhus to use what is probably the best wind tunnel facilities for
studying aeolian sand transport in the world.

Among the early experiments in the wind tunnel was an experiment to
determine how the distance travelled by a sand grain depends on its
size. A ripple of blue coloured sand was placed at the upwind end of
the tunnel. After 2.5 minutes of wind transport, samples of the
uppermost layer of the sand bed were taken at nine location along the tunnel by 
spraying the surface with an adhesive spray. Each sand sample
was separated by sieving into five size classes, and in each of
these the fraction of coloured sand was determined. By a statistical
analysis of these data, the mean and standard deviation of the transport
speed was determined for each size fraction; for details see
\cite{oebn82d}. This population approach was later supplemented by an
individual approach where sand grains in a few size classes were
labelled with radioactive gold, see \cite{oebn85b}. In each run a
single radioactive grain was followed with a scintillation detector,
so  the data contained much more information about the motion of the
grains, which turned out to be well described by a compound Poisson
process with exponentially distributed jumps that is sometimes
interrupted while the grain is buried, presumably under a ripple.  

The 1980s was a period of swift international progress in the
scientific understanding of
aeolian sediment transport, see \cite{ms91}, where the Sand Gang
played a prominent role. During the 1990s Barndorff-Nielsen's focus
turned to financial econome\-trics, see \cite{Shephard}, but members of
the Sand Gang continued to publish influential papers such 
as \cite{ms04}, which presented a formula relating the transport rate
of sand to the wind speed based on theoretical work and experiments
in the wind tunnel. A longstanding objective for the Sand Gang was
a mathematical model that could explain why the log-size
distribution of natural sand is often hyperbolic. \cite{oebn77a}
had proposed that the representation of the
hyperbolic distribution as a variance-mean mixture might be used for
this, and in \cite{ms} a model of this type was established. A
breakage model that would in itself imply that the logarithm of the
grain size is $N(\mu + \beta \tau, \tau)$ distributed, is modified by
taking into account that the travel time $\tau$ between source and
deposit varies randomly between the individual grains. If the motion
of a grain is approximated by a Brownian motion with drift, the first
arrival time $\tau$ is inverse Gaussian distributed, and hence the
log-size is NIG-distributed. \citet[p.\ 142]{oebn18a} concluded that
``this result, in a sense, is the ultimate theoretical justification
for the application of the hyperbolic and normal inverse Gaussian laws
in sedimentology''. 

\subsection{Hyperbolic processes and turbulence}

Right from the start, Barndorff-Nielsen was interested in stochastic
processes with hyperbolic marginal distributions, mainly motivated by
the search for explanations of why the hyperbolic 
distribution and its generalizations fit data from so many areas of research. Already in 
\cite{oebn78a} he proposed the {\it hyperbolic diffusion process} given as
the solution to the stochastic differential equation
\begin{equation} \label{hypdiff}
  dX_t = \mbox{\small{$\frac{1}{2}$}}\sigma^2\left[\beta-\alpha\frac{X_t-\mu}{
  \sqrt{\delta^2+(X_t-\mu)^2}}\right]dt + \sigma dW_t.
\end{equation}
This is the one-dimensional Langevin diffusion with constant diffusion
coefficient and with the $H(\alpha, \beta, \delta, \mu)$ distribution as
its stationary distribution. Barndorff-Nielsen later turned his attention to
L\'evy processes and L\'evy-driven processes and
did not study hyperbolic diffusion  processes further, but several
such processes are presented in \cite{bibbysorensen}. 
Barndorff-Nielsen's contributions to classical and free infinite
divisibility and L\'evy processes are reviewed in \cite{VictorSteen}.
The generalized hyperbolic distributions are infinitely divisible, see
\cite{oebn77b}, so for any given generalized hyperbolic distribution,
a {\it L\'evy process} $X$ exists such that $X_1$ has this given distribution.
This is particularly useful for the normal inverse Gaussian and the
variance gamma distributions, where the convolution results
(\ref{nigconvolution}) and (\ref{vgconvolution}) imply that the
distribution of $X_t$ is NIG or variance gamma for all $t$. The NIG
L\'evy process was studied thoroughly in \cite{oebn97a}. 

In the 1980s, Barndorff-Nielsen's motivation for developing hyperbolic
stochastic processes came mainly from his interest in {\it turbulence}. Already
around 1980, he had the vision that the closure problem in the theory 
of turbulence could be solved, or circumvented, by developing a
suitable parametric stochastic process model for the turbulent
velocity field. This approach served him well in his life-long 
efforts to obtain a deeper understanding of the phenomenon of
turbulence. Since the 1920s, the main approach to solving the problem
of turbulence had been to attempt to determine (mixed) moments of all
orders of the turbulent field via differential equations for the moments obtained
from the equations of motion for fluid flows, mainly the Navier-Stokes
equation. Unfortunately, as a consequence of the non-linearity of the
equations of motion, it is impossible to obtain a closed system of
equations for a finite number of moments - the number of unknowns will
always be larger than the number of equations. At the time when
Barndorff-Nielsen became interested in turbulence, most theoretical
work on the dynamics of turbulence was still devoted to overcoming
this closure problem, so new ideas were required. More details on the
development of the theory of turbulence up to this time can be found
in the introduction to \cite{moninyaglom}. Some first suggestions
of Barndorff-Nielsen's approach were made in \cite{oebn79a}, and the
approach was more fully discussed in \cite{oebn90a}.

In \cite{oebn79a} data sets consisting of velocity differences (over
space and time) measured in three different kinds of turbulent fields were
analysed. It was established that the distribution of the velocity
differences were to a good approximation hyperbolic, except that the
tails tended to be heavier than the tails of a hyperbolic
distribution. These findings were confirmed in \cite{oebn04a}, where it
was found that NIG-distributions provide an excellent fit to the
distributions of the velocity differences. An application of the
NIG-version of the shape triangle showed how the distributions
converge to the normal distribution as the space or time difference
becomes large. Data on wind shear, which constitute a major problem
for air traffic during landing, was shown to be modelled well and
parsimoniously by hyperbolic distributions in \cite{oebn89a}.

Some first steps towards a parametric statistical model for the velocity and
velocity derivative fields in stationary turbulence were taken in
\cite{oebn90a}. Based on experimental evidence, including the findings in
\cite{oebn79a}, and theoretical results concerning the statistical
properties of the velocity and velocity difference processes in the
mean wind direction of a stationary, high Raynolds number turbulent
wind field, five desiderata for constructing a parametric stochastic 
process model were formulated. In an attempt to meet these desiderata,
a number of stochastic process models with marginal distributions of
hyperbolic type were proposed. Most notably, processes defined as the
sum of $m$ independent stationary AR(1)-processes $X^{(i)}_n = \rho_i
X^{(i)}_{n-1} + \epsilon^{(i)}_n$ were considered. The lag $k$ auto-correlation of
such a process has the form $\sum_{i=1}^m \phi_i \rho_i^k$, where
$\phi_1 + \cdots + \phi_m = 1$.  It was shown that
a stationary process of this type with invariant distribution $D$
exists if and only if $D$ is self-decomposable, and the characteristic
functions were found of the distributions of the innovations $\epsilon^{(i)}_n$
needed to obtain a given invariant distribution and given values of
$\rho_i$ and $\phi_i$. The generalized hyperbolic distributions are
self-decomposable, see \cite{halgreen}, and can thus be the marginal
distribution of a process of the proposed type. The generalized 
logistic distributions are self-decomposable too, see \cite{oebn82c},
and manageable expressions for the distributions of the innovation
processes were found for 
symmetric generalized logistic distributions. The corresponding
process with $m=2$ was fitted to data from a field experiment carried out at
Ferring Beach on the Danish west coast in 1985. The process with $m=2$
fits the auto-correlation function well: the largest regression
coefficient describes the correlation over long ranges, while the
other modifies the correlation structure at short ranges. The data
were further analysed in \cite{oebn93a}, where it was found that the
logarithm of the spectral density of the estimated process is well
approximated by a straight line with slope -5/3 in the so-called
intertial sub-range, which is in agreement with Kolmogorov's
5/3-law for locally isotropic turbulence.

A thorough mathematical study of the processes proposed and applied in
\cite{oebn90a, oebn93a} was presented in
\cite{oebn98b}. Moreover, a continuous time version of the sums of
AR(1) processes was presented and investigated, namely processes that
are sums of independent, stationary Ornstein-Uhlenbeck processes
driven by L\'evy processes. Specifically, $Y_t = X^{(i)}_t + \cdots +
X^{(m)}_t$, where $X^{(i)}, \ldots, X^{(m)}$ are independent
processes given by
\[
X_t^{(i)} = e^{-\gamma_i t} X_0^{(i)} + \int_0^t e^{-\gamma_i
  (t-s)}dZ_s^{(i)}, \ \ \gamma_i > 0. 
\]
Here $Z^{(i)}$ is a L\'evy process independent of $ X^{(i)}_0$. Also
in the continuous time case, a stationary process of this type with
invariant distribution $D$ exists if and only if $D$ is
self-decomposable. The L\'evy symbols of the processes $Z^{(i)}$ were
found, for which a given invariant distribution and a given
autocorrelation function $\sum_{i=1}^m \delta_i e^{-k \gamma_i}$ with
$\delta_1 + \cdots + \delta_m = 1$ is obtained. \cite{oebn98a}
introduced the name {\it background driving L\'evy processes} for the
processes $Z^{(i)}$. Importantly, he determined the background
driving L\'evy processes for which the invariant distribution of $Y_t$
is a NIG distribution as a sum of three independent L\'evy processes:
a NIG L\'evy process, a compound Poisson process and a third process
for which a simple expression for the L\'evy measure was
found. Similarly, it was shown that the background driving L\'evy processes
needed to obtain an inverse Gaussian invariant distribution are sums of
two independent L\'evy processes: an inverse Gaussian L\'evy process
and a compound Poisson process.

At this time Barndorff-Nielsen had become deeply interested in
financial econometrics, and in \cite{oebn98a} he discussed common
characteristic features of observational time series from
turbulence and finance, including semi-heavy tails, asymmetry,
varying activity (called volatility in finance and intermittency in
turbulence), and aggregational Gaussianity. In the next many years the
cross fertilisation between these two field played a crucial role in
Barndorff-Nielsen's research, where modelling ideas from one field turned out
to be well suited for the other. A first and important instance
was the model for the price of a financial asset presented in
\cite{oebn01b}, now known as the Barndorff-Nielsen/Shephard (or BNS)
model, where the volatility is 
modelled by a sum of  independent, stationary L\'evy-driven
Ornstein-Uhlenbeck processes. Barndorff-Nielsen's many contributions
to financial econometrics are reviewed in \cite{Shephard}.

Already while working on \cite{oebn90a}, Barndorff-Nielsen felt that
it would be worthwhile to generalize the sums of Ornstein-Uhlenbeck
processes to intergrals of such processes with respect to a suitable
measure. He did this in \cite{oebn01a}, where he introduced the class
of superpositions of Ornstein-Uhlenbeck type processes, which turned
out to be very useful in both finance and turbulence. In the following
years Barndorff-Nielsen's research on turbulence modelling developed
dramatically in close collaboration with J\"urgen Schmiegel. A solid
basis was obtained in the three papers \cite{oebn04b, oebn09a} and
\cite{oebn04a} using temporal-spatial models defined as
integrals with respect to
a L\'evy basis, including superpositions of Ornstein-Uhlenbeck
processes. In the third of these papers the Brownian semistationary
processes were introduced. Through a number of steps the project
over the years produced very sophisticated models of turbulence and for
other interesting applications. As a crucial part of this process, the
{\it ambit processes} were developed as a natural modelling framework for
turbulence modelling, and ambit stochastics became a new branch of
probability theory. Barndorff-Nielsen's research on ambit stochastics
is reviewed in \cite{Veraart}. Modelling of turbulence by ambit
processes represented a break away from the then prevailing attempts
to model turbulence by means of multifractality, but interestingly \cite{oebn05a}
demonstrated that the L\'evy-based approach encompasses aspects of
multifractality. As Barndorff-Nielsen liked to emphasize, the ambit
approach to understanding the nature of turbulence is fundamentally
different from the traditional approach via 
the Navier-Stokes equation driven by random forces. It is therefore
interesting, that \cite{birnir} showed that for a certain choice of the stochastic
forcing of the Navier-Stokes equation, the velocity increments in
the longitudinal direction are exactly NIG-distributed.

\end{document}